\documentclass[a4paper,11pt]{article}
\usepackage{amsfonts}
\usepackage{amsmath}
\usepackage{epsfig}
\usepackage{latexsym}
\usepackage{amssymb}
\usepackage{color}
\usepackage{amscd}
\usepackage{multirow}
\usepackage{graphicx}
\usepackage{lscape}
\usepackage{bm}
\usepackage[ruled,vlined]{algorithm2e}

\def\R{{\mathbb R}}

\newcommand{\Remm}[1]{}

\newtheorem{model ass}[theo]{Model Assumptions}

\numberwithin{equation}{section}

\begin{document}

\noindent{\Large {\textbf{Model Selection and Adaptive Markov
chain Monte Carlo for Bayesian Cointegrated VAR model}}}

\vspace{1cm} \noindent\textbf{Gareth W.~Peters}\newline
{\footnotesize {Lecturer, Department of Mathematics and Statistics, 
University of New South Wales, Sydney, 2052, Australia;\newline
CSIRO Mathematical and Information Sciences, Locked Bag 17,
North Ryde, NSW, 1670, Australia;\newline
email: garethpeters@unsw.edu.au \newline
{\textit{Corresponding author.}}}}

\vspace{0.5cm} \noindent\textbf{Balakrishnan ~Kannan} \newline
{\footnotesize Baronia Capital Pty. Ltd., 12 Holtermann St., Crows
Nest, NSW 2065, Australia.\newline
e-mail:Balakrishnan.Kannan@baroniacapital.com.au}

\vspace{0.5cm} \noindent\textbf{Ben ~Lasscock} \newline
{\footnotesize Baronia Capital Pty. Ltd., 12 Holtermann St., Crows Nest, NSW 2065, Australia.\newline
e-mail:ben.lasscock@baroniacapital.com.au}

\vspace{0.5cm} \noindent\textbf{Chris ~Mellen} \newline
{\footnotesize Baronia Capital Pty. Ltd., 12 Holtermann St., Crows Nest, NSW 2065, Australia.\newline
e-mail: Chris.Mellen@baroniacapital.com.au}


\begin{center}
\textbf{Submitted: 16 July 2009\\
Revision 1: 10 November 2009\\
Revision 2: 20 April 2010}
\end{center}
\newpage
\begin{abstract}


\noindent This paper develops a matrix-variate adaptive Markov chain Monte Carlo (MCMC) methodology for Bayesian Cointegrated Vector Auto Regressions (CVAR). We replace the popular approach to sampling Bayesian CVAR models, involving griddy Gibbs, with an automated efficient alternative, based on the Adaptive Metropolis algorithm of Roberts and Rosenthal, (2009). Developing the adaptive MCMC framework for Bayesian CVAR models allows for efficient estimation of posterior parameters in significantly higher dimensional CVAR series than previously possible with existing griddy Gibbs samplers. For a n-dimensional CVAR series, the matrix-variate posterior is in dimension $3n^2 + n$, with significant correlation present between the blocks of matrix random variables. Hence, utilizing a griddy Gibbs sampler for large n becomes computationally impractical as it involves approximating an $n \times n$ full conditional posterior using a spline over a high dimensional $n \times n$ grid. The adaptive MCMC approach is demonstrated to be ideally suited to learning on-line a proposal to reflect the posterior correlation structure, therefore improving the computational efficiency of the sampler.

We also treat the rank of the CVAR model as a random variable and perform joint inference on the rank and model parameters. This is achieved with a Bayesian posterior distribution defined over both the rank and the CVAR model parameters, and inference is made via Bayes Factor analysis of rank.

Practically the adaptive sampler also aids in the development of automated Bayesian cointegration models for algorithmic trading systems considering instruments made up of several assets, such as currency baskets. Previously the literature on financial applications of CVAR trading models typically only considers pairs trading (n=2) due to the computational cost of the griddy Gibbs. We are able to extend under our adaptive framework to $n >> 2$ and demonstrate an example with n = 10, resulting in a posterior distribution with parameters up to dimension 310. By also considering the rank as a random quantity we can ensure our resulting trading models are able to adjust to potentially time varying market conditions in a coherent statistical framework.



\vspace{0.5cm} \noindent \textbf{Keywords:} Cointegrated Vector Auto Regression,
Adaptive Markov chain Monte Carlo, Bayesian Inference, Bayes Factor.
\end{abstract}

\pagebreak
\section{Introduction}

Bayesian analysis of Cointegrated Vector Auto Regression (CVAR) models has been addressed in several papers, see Koop {\textit{et al.}} (2006) for an overview. In a Bayesian CVAR model, specification of the matrix-variate model parameters priors, to ensure the posterior is not improper, must be done with care, see Koop {\textit{et al.}} (2006). This has significant implications on the Bayesian model structure, in particular one can not make a blind specification of priors on the VAR model coefficients as it may result in improper posterior distributions. For this reason it is common to consider the Error Correction Model (ECM) framework, see for example p.141-142 of Reinsel and Velu (1998). In this paper we do not aim to address the issue of prior choice or prior distortions and we adopt the model of Sugita (2002) and Geweke (1996) which admits desirable conjugacy properties. The resulting posterior for a n-dimensional CVAR series, is matrix-variate in dimension up to $3n^2 + n$ for full rank models, with significant correlation present between and within the blocks of matrix random variables. This presents a challenge to efficiently sample from the posterior distribution when n is large. 

The focus of the paper and novelty introduced involves developing a Bayesian adaptive MCMC sampling, based on the proposed algorithm of Roberts and Rosenthal, (2009), to allow us to significantly increase the dimension, n, of the CVAR series that can be estimated. Typically in the cointegration literature the sampling approach adopted is a griddy Gibbs sampling framework, see Bauwens and Lubrano (1996), Bauwens and Giot (1997), Geweke (1996), Kleibergen and van Dijk (1994), Sugita (2002) and Sugita (2009).  The conjugacy properties of the Bayesian model we consider result in exact sampling of two of the matrix-variate random variables corresponding to the unknown error covariance matrix and the combined matrix random variable containing the cointegration equilibrium reversion rates $\bm{\alpha}$ and the mean level $\bm{\mu}$ of the CVAR series. However, the third unknown matrix-variate random variable corresponding to the cointegration vectors $\bm{\beta}$ has a marginal posterior distribution with support in dimension $n \times r$. When the cointegration rank $r$ and the dimension of the CVAR series $n$ is large ($n > 5$) then the standard griddy Gibbs based samplers are no longer computationally viable samplers. Alternative samplers which may attempt to deconstruct the full conditional distribution of the posterior for the cointegration vectors $\bm{\beta}$ into components of this matrix, updating them one at a time will run into significant difficulties with efficiency in the mixing properties of the resulting Markov chain. The reason for this is due directly to two factors: the identification normalization constraint of the matrix $\bm{\beta}$; and the strong correlation present in the full conditional posterior distribution for the matrix random variable $\bm{\beta}$. Hence, utilizing a griddy Gibbs sampler for large $n$ becomes computationally impractical as it involves approximating up to an $n \times n$ matrix-variate full conditional posterior using a spline constructed over a high dimensional space with d knot points per dimension, creating a requirement for $d^n$ total grid points. The sampler we develop overcomes these difficulties utilizing an adaptive MCMC approach. We demonstrate that it is ideally suited to learning on-line a proposal to reflect the posterior correlation in the matrix-variate random variable, ensuring that updating this $n \times r$ matrix at each stage of the adaptive MCMC algorithm results in a non-trivial acceptance probability.

Adaptive MCMC is a new methodology to learn on-line the 'optimal' proposal distribution for an MCMC algorithm, see Atachade and Rosenthal (2005), Haario, Saksman and Tamminen (2001; 2007) and Andrieu and Moulines (2006) and more recently Giordani and Kohn, (2006) and Silva \textit{et al.}, (2009), of which there are several different versions of adaptive MCMC and Particle MCMC algorithms. Basically adaptive MCMC algorithms aim to allow the Markov chain to adapt the Markov proposal distribution online throughout the simulation in such a way that the correct stationary distribution is still preserved, even though the Markov transition kernel of the chain is changing throughout the simulation. Clearly, this requires careful constraints on the type of adaption mechanism and the adaption rate to ensure that stationarity is preserved for the resulting Markov chain. 

To summarize, this paper extends the matrix-variate block Gibbs sampling framework typically used in Bayesian Cointegration models by replacing the computational $n \times n$ dimensional griddy Gibbs sampler with two possible automated alternatives which are based on matrix-variate adaptive Metropolis-within-Gibbs samplers. Additionally, we consider rank estimation for reduced rank Cointegration models. From a Bayesian perspective we tackle this via Bayes Factor (BF) analysis for posterior "model" probabilities of the rank. Then we demonstrate estimation and predictive performance under a Bayesian setting for both Bayesian Model Selection (BMS) and Bayesian Model Averaging (BMA).

The models and algorithms developed allow for estimation of either the rank $r$, i.e. the model index, and the lag $p$ of the CVAR model jointly with the model parameters. For simplicity we shall assume the lag is fixed and known.

In this paper the following notation will be used: ' denotes
transpose, $I_d$ is the $d \times d$ identity matrix, p(.) denotes a density and P(.) a distribution,
$\Omega$ will be the space on which densities will take their
support and it will be assumed throughout that we are working with
Lebesgue measure. The operator $\otimes$ denotes the Kronecker
product, $\|\cdot\|$ denotes the total variation norm and $\triangle$ denotes the unit vector difference operator.
We denote generically the state of a Markov chain at time $j$ by random
variable $\Theta^{(j)}$ and the transition kernel from realized
state $\Theta^{(j-1)}=\theta$ to $\Theta^{(j)}$ by $Q\left(\theta
, \Theta^{(j)}\right)$. In the case of an adaptive transition
kernel we will also assume that there is a sequence of transition
kernels denoted by $Q_{\Gamma_j}\left(\theta ,
\Theta^{(j)}\right)$, where $\Gamma_j$ is the sequence index.

\subsection{Contribution and structure}

In section \ref{modelCVAR} we present the matrix-variate posterior distribution for the CVAR model formulated under an Error Correction Model (ECM) model framework. Next in section \ref{modelCVAR2} we discuss the Bayesian CVAR model conditional on knowledge of the co-integration rank. This includes discussing and summarizing properties of the Bayesian CVAR model including identification, the justification of the ECM framework and issues to consider when selecting matrix priors for Bayesian CVAR models with respect to prior distortions. At this stage we make explicit the justification for why the Bayesian model decomposes the cointegration matrix $\Pi = \bm{\alpha}\bm{\beta}'$ under the ECM framework, since working directly with $\Pi$ precludes direct use of Monte Carlo samples for inference in the VAR model setting.  As pointed out in Geweke (1996) and Sugita (2002), conditional on matrix $\bm{\beta}$ the nonlinear ECM model becomes linear and therefore under the informative priors we utilize, we can once again apply standard Bayesian analysis to the VAR model, this turns out to be a very useful property widely used in the cointegration literature.
Then in section \ref{modelCVAR3} we present the two algorithms developed based on Adaptive MCMC to obtain samples from the target posterior, followed by section \ref{modelCVAR4} which presents the framework for rank estimation we utilize, along with discussion of model selection and model averaging, with respect to the unknown rank of the CVAR system. We conclude with both synthetic simulation examples with n ranging from 4 to 10, resulting in posteriors defined in dimensions between 52 and 310 dimensions. We also provide analysis on two real data examples from pairs and triples trading typically considered in real world financial algorithmic trading models.

\section{CVAR model under ECM framework}
\label{modelCVAR}

We note that a well presented representation to co-integration models is provided by Engle and Granger (1987), Sugita, (2002), Sugita, (2009) and for the original error correction representation of a co-integrated series, see Granger (1981) and Granger and Weiss (1983). The model presented in Sugita, (2009) is based on the model of Strachen and van Dijk, (2007) and it generalizes the VECM model in Sugita, (2002) to include explicitly the possibility of an intercept and a linear time trend. In this paper we will consider a CVAR model in which we have an intercept term but no time trend, the extension to include a time trend is trivial to incorporate into our simulation methodology. Based on the definitions of Sugita, (2002) for a co-integrated series, we denote the vector observation at time $t$ by $\bm{x}_{t}$. Furthermore, we assume $\bm{x}_{t}$ is an integrated of order 1, I(1), $(n \times 1)$-dimensional vector with $r$ linear cointegrating relationships. The error vector at time $t$, $\bm{\epsilon}_t$ are assumed time independent and zero mean multivariate Gaussian distributed, with covariance $\Sigma$. The Error Correction Model (ECM) representation we consider is given by,
\begin{equation}
\label{ECM1}
\triangle \bm{x}_t = \bm{\mu} + \bm{\alpha}\bm{\beta}'\bm{x}_{t-1} + \sum_{i=1}^{p-1}\Psi_i \triangle \bm{x}_{t-i} + \bm{\epsilon}_t
\end{equation}
where $t = p, p+1, \ldots, T$ and $p$ is the number of lags. Furthermore, the matrix dimensions are: $\bm{\mu}$ and $\bm{\epsilon}_t$ are $(n \times 1)$, $\Psi_i$ and $\Sigma$ are $(n \times n)$, $\bm{\alpha}$ and $\bm{\beta}$ are $(n \times r)$.  

We can now re-express the model in equation (\ref{ECM1}) in a multivariate regression format, as follows
\begin{equation}
    Y = X\Gamma + Z\bm{\beta}\bm{\alpha}' + E = WB + E,
\label{eqn:1}
\end{equation}
where,
\begin{align*}
\mbox{$Y$} =
\left(
\begin{array}{cccc}
\triangle \bm{x}_{p} & \triangle \bm{x}_{p+1} & \ldots & \triangle \bm{x}_{T} \end{array}
\right) ',
\mbox{$Z$} =
\left(
\begin{array}{cccc}
\bm{x}_{p-1} & \bm{x}_{p} & \ldots & \bm{x}_{T-1} \end{array}
\right) ' \\
\mbox{$E$} =
\left(
\begin{array}{cccc}
\bm{\epsilon}_{p} & \bm{\epsilon}_{p+1} & \ldots & \bm{\epsilon}_{T} \end{array}
\right) ',
\mbox{$\Gamma$} =
\left(
\begin{array}{cccc}
\bm{\mu} & \Psi_{1} & \ldots & \Psi_{p-1} \end{array}
\right) ' \\
\mbox{$X$} =
\left(
\begin{array}{cccc}
1 & \triangle \bm{x}_{p-1}' & \ldots & \triangle \bm{x}_{1}' \\
1 & \triangle \bm{x}_{p}' & \ldots & \triangle \bm{x}_{2}' \\
\vdots & \vdots & \ldots & \vdots \\
1 & \triangle \bm{x}_{T-1}' & \ldots & \triangle \bm{x}_{T-p+1}' \end{array}
\right),
\mbox{$W$} =
\left(
\begin{array}{cc}
X & Z\bm{\beta} \end{array}
\right),
\mbox{$B$} =
\left(
\begin{array}{cc}
\Gamma' & \bm{\alpha} \end{array}
\right) 
\end{align*}
Here, we let $t$ be the number of rows of $Y$, hence $t = T-p+1$, producing $X$ with dimension $t \times (1+n(p-1))$, $\Gamma$ with dimension $((1+n(p-1)) \times n)$, $W$ with dimension $t \times k$ and $B$ with dimension $(k \times n)$, where $k = 1+n(p-1)+r$, see Sugita, (2002) for additional details regarding this parameterization.
The parameters $\mu$ represents the trend coefficients, and $\Psi_{i}$ is the $i^{th}$ matrix of
autoregressive coefficients and the long run multiplier matrix is
given by $\Pi = \bm{\alpha}\bm{\beta}'$. 

The long run multiplier matrix is an important quantity of this model, its properties include: if
$\Pi$ is a zero matrix, the series $\bm{x}_t$ contains n unit roots; if
$\Pi$ has full rank then each univariate series in $\bm{x}_t$ are
(trend-)stationary; and co-integration occurs when $\Pi$ is of
rank $r<n$. The matrix $\bm{\beta}$ contains the co-integration
vectors, reflecting the stationary long run relationships between
the univariate series within $\bm{x}_t$ and the $\bm{\alpha}$ matrix contains
the adjustment parameters, specifying the speed of adjustment to
equilibria $\bm{\beta}'\bm{x}_t$.

This results in a likelihood model, where the parameters of interest are $B$, $\Sigma$ and $\bm{\beta}$, given by
\begin{equation*}
\begin{split}
\mathcal{L}(B,\Sigma,\bm{\beta}|Y) &= (2\pi)^{-0.5nt}|\Sigma \otimes I_t|^{-0.5} \exp\left(-0.5 Vec(Y-WB)'(\Sigma^{-1}\otimes I_t^{-1})Vec(Y-WB)\right) \\
& \propto |\Sigma|^{-0.5t}\exp\left(-0.5 tr[\Sigma^{-1}(\hat{S} + R)]\right),
\end{split}
\end{equation*}
where $\Sigma = Cov(E)$ and
\begin{align*}
R = (B-\hat{B})'W'W(B-\hat{B}),
\hat{S} = (Y-W\hat{B})'(Y-W\hat{B}),
\hat{B} = (W'W)^{-1}W'Y
\end{align*}

\section{Bayesian CVAR models conditional on Rank (r)}
\label{modelCVAR2}

The assumptions and restrictions of our Bayesian CVAR model include:
\begin{enumerate}
\item{ {\textbf{Identification Issue:}} For any non-singular matrix $A$, the matrix of long run multipliers $\Pi = \bm{\alpha}\bm{\beta}'$ is indistinguishable from $\Pi = \bm{\alpha} A A^{-1} \bm{\beta}'$, see Koop \textit{et al.} (2006) or Reinsel and Velu (1998). We use a standard approach to globally overcome this problem by incorporating a non unique identification constraint. We impose $r^2$ restrictions as follows $\bm{\beta} = [I_{r}, \bm{\beta}_{*}']'$, where $I_{r}$ denotes the $r \times r$ identity matrix. However, as noted by Kleibergen and van Dijk (1994) and discussed in Koop \textit{et al.} (2006) this can still result in local identification issues at the point $\bm{\alpha}=\bm{0}$, when $\bm{\beta}$ does not enter the model. Hence, one must be careful to ensure that the Markov chain generated by the matrix-variate block Gibbs sampler is not invalidated by the terminal absorbing state. As is standard we monitor the performance of the sampler to ensure this has not occurred.}

\item{{\textbf{Error Correction Model:}} The ECM framework complicates Bayesian analysis since products, $\bm{\alpha}\bm{\beta}'$, preclude direct use of Monte Carlo samples for inference in the VAR model setting. However, conditional on $\bm{\beta}$ the nonlinear ECM model becomes linear and therefore under the informative priors used by Geweke (1996) and Sugita (2002), we can once again apply standard Bayesian analysis to the VAR model.}

\item{{\textbf{Prior Choices:}} We do not consider the issue of prior distortions illustrated by Kleibergen and van Dijk (1994). This is not the focus of the present paper. Alternative prior models in the cointegration setting include Jeffrey's priors, Embedding approach and a focus on the cointegration space.}
\end{enumerate}

\subsection{Prior and Posterior Model}
Here we present the model for estimation of $\bm{\beta}$, $B$ and
$\Sigma$ conditional the rank $r$. As in Sugita (2002), we use a
conjugate hierarchical prior.
\begin{itemize}

    \item $\bm{\beta} \sim  N(\bar{\bm{\beta}},Q \otimes H^{-1})$ where $N(\bar{\bm{\beta}},Q \otimes H^{-1})$ is the matrix-variate Gaussian distribution with prior mean $\bar{\bm{\beta}}$, Q  is an $(r \times r)$ positive definite matrix, H an $(n \times n)$ matrix.

    \item $\Sigma \sim IW(S,h)$ where $IW(S,h)$ is the Inverse Wishart distribution with h degrees of freedom and S is an $(n \times n)$ positive  definite matrix.

    \item $B|\Sigma \sim N(P,\Sigma \otimes A^{-1} )$ where $N(P,\Sigma \otimes A^{-1})$ is the matrix-variate Gaussian distribution with prior mean $P$ which is $k \times n$ and $A$ is a $(k \times k)$ matrix, with $k = n(p-1) + 1 + r$ which corresponds to the number of columns in $W$.

\end{itemize}
Combining the priors and likelihood produce matrix-variate conditional posterior distributions (derivation details provided in Sugita, (2002)):
\begin{itemize}
\item{ Inverse Wishart distribution for $p(\Sigma|\bm{\beta},Y) \propto |S_{\star}|^{(t+h)/2}|\Sigma|^{-(t+h+n+1)/2}\exp\left(-0.5 tr(\Sigma^{-1}S_{\star})\right)$ which is trivial to sample exactly;}
\item{Matrix-variate Gaussian for $p(B|\bm{\beta},\Sigma,Y) \propto |A_{\star}|^{n/2}|\Sigma|^{-k/2}\exp\left(-0.5 tr\left(\Sigma^{-1}(B-B_{\star})'A_{\star}(B-B_{\star})\right)\right)$ 
(or alternatively matrix-variate student-t distribution form for $p(B|\bm{\beta},Y)$), both trivial to sample exactly;} 
\item {The marginal matrix-variate posterior for the cointegration vectors, $\bm{\beta}|Y$, is not well studied and is given by
\label{eqn:fcdbeta}
\begin{equation}
p(\bm{\beta}|Y) \propto p(\bm{\beta})|S_{*}|^{-(t+h+1)/2}|A_{*}|^{-n/2}.
\end{equation} }
\end{itemize}
where we define $A_{\star} = A + W'W$, $B_{\star}=(A+W'W)^{-1}(AP+W'W\widehat{B})$ and $S_{\star}=S+\widehat{S}+(P-\widehat{B})'[A^{-1}+(W'W)^{-1}]^{-1}(P-\widehat{B})$.

\section{Sampling and Estimation Conditional on Rank r}
\label{modelCVAR3}

Here we focus on obtaining samples from the posterior distribution which can be used to obtain Bayesian parameter estimates (MMSE, MAP). The complication in sampling arises with the full conditional posterior \ref{eqn:fcdbeta} which can not be sampled from via straight forward inversion sampling.

In this paper we outline novel algorithms to sample from the posterior distribution \newline $p\left(\bm{\beta}|Y,B,\Sigma,r\right)$, providing an alternative automated approach to the griddy Gibbs sampler algorithm made popular in this Bayesian co-integration setting by Bauwens and Lubrano (1996).

The matrix-variate griddy Gibbs sampler numerically approximates
the target posterior on a grid of values and then performs
numerical inversion to obtain samples from \ref{eqn:fcdbeta} at
each stage of the MCMC algorithm. Such a grid based procedure will
suffer from the curse of dimensionality when $n$ is large $\left(n
> 5\right)$ after which it becomes highly inefficient. Note, alternative approaches such as Importance Sampling will also
be problematic once $n$ becomes too large. It is difficult to
optimize the choice of the Importance Sampling distribution which
will minimize the variance in the importance weights.

Instead we propose alternative samplers using adaptive matrix-variate MCMC methodology. They do not suffer from the curse of dimensionality and are simple to implement and automate.

\begin{itemize}
\item{ \textbf{Algorithm 1 - Random Walk (mixture local \& global moves):} Involves an offline adaptively pretuned mixture proposal containing a combination of local and global Random Walk (RW) moves. The proposal for the local RW moves have standard deviation tuned to produce average acceptance probabilities between [0.3, 0.5]. The independent global matrix-variate proposal updates all elements of $\bm{\beta}$ via a multivariate Gaussian proposal centered on Maximum Likelihood parameter estimates for $\bm{\beta}$ and the Fisher information matrix for the covariance of the global proposal. This is similar to the approach adopted in Vermaak {\textit{et al.}} (2004) and Fan {\textit{et al.}} (2009).}
\item{ \textbf{Algorithm 2 - Adaptive Random Walk:} Involves an online matrix-variate adaptive Metropolis algorithm based on methodology presented in Roberts and Rosenthal (2009). }
\end{itemize}

Proceeding sections denote the algorithmic 'time' index by $j$ and the current state of a Markov chain for generic parameter $\theta$ at time $j$ by $\theta^{(j)}$. The length of the Markov chain is $J$.

Note, since we have imposed $r^2$ restrictions in the form of $I_{r}$, any proposal for $\bm{\beta} = [I_r, \tilde{\bm{\beta}}]$ will only correspond to the unrestricted elements of $\bm{\beta}$ denoted by $\tilde{\bm{\beta}}$. In our case, these correspond to those in locations $\left(n-r\right) \times r$.

\subsection{Algorithm 1}
In Algorithm 1 the mixture proposal distribution for parameters $\tilde{\bm{\beta}}$ will be given by,
\begin{equation}
q\left(\tilde{\bm{\beta}}^{(t-1)},\cdot\right)= w_1 N\left(\tilde{\bm{\beta}};\tilde{\bm{\beta}}^{ML},\Sigma^{ML}\right) + \left(1-w_1\right) \prod_{i=1}^{(n-r)\times r} N\left(\tilde{\bm{\beta}}_{i,k};\tilde{\bm{\beta}}^{(t-1)}_{i,k},\sigma^2_{i,k}\right).
\end{equation}
The Maximum Likelihood parameters are obtained off-line, see (p.
286 Lutkepohl (2007)). The local random walk proposal variances
$\sigma^2_{i,k}$ for each element of $\tilde{\bm{\beta}}$ are obtained
via pre-tuning.

{\small{
\begin{algorithm}
\dontprintsemicolon
\KwIn{ Initial Markov chain state $\left(\Sigma^{(0)}, B^{(0)}, \bm{\beta}^{(0)}\right)$. }
\KwOut{Markov chain samples
$\{\Sigma^{(j)},B^{(j)},\bm{\beta}^{(j)}\}_{j=1:J} \sim
p\left(\Sigma,B,\bm{\beta}|Y\right)$.}
\Begin{
\begin{enumerate}
\item Set initial state $\left(\Sigma^{(0)}, B^{(0)}, \bm{\beta}^{(0)}\right)$ deterministically or by sampling the priors.\;
\item Calculate Maximum Likelihood parameters $\tilde{\bm{\beta}}^{ML}$ and $\Sigma^{ML}$.\;
\item Initialize $w_1$ and $w_2 = 1-w_1$ and index $j=1$.\;
\end{enumerate}
\Repeat{j = J}{
\begin{enumerate}
\item[5.] Sample $\Sigma$ via inversion to obtain $\Sigma^{(j)}$.\;
\item[6.] Sample $B$ via inversion to obtain $B^{(j)}$.\;
\item[7.] Sample realization $U=u$ where $U\sim U[0,1]$\;
\end{enumerate}
\eIf(\tcc*[f]{perform a local random walk move}){$u \geq w_1$}{
\begin{enumerate}
\item[7a.] Sample uniformly index $(i,k)$ from set of $n-r \times r$ elements.\;
\item[7b.] Sample the $(i,k)$-th component $\tilde{\bm{\beta}}^*_{i,k} \sim N\left(\tilde{\bm{\beta}}_{i,k};\tilde{\bm{\beta}}^{(j-1)}_{i,k},\sigma^2_{i,k}\right)$.\;
\item[7c.] Construct proposal $\bm{\beta}^* = [I_{r \times r},\tilde{\bm{\beta}}^*]$, where $\tilde{\bm{\beta}}^*$ is $\tilde{\bm{\beta}}^{(j-1)}$ with \\ the $(i,k)$-th element given by $\tilde{\bm{\beta}}^*_{i,k}$.\;
\end{enumerate}
}(\tcc*[f]{perform a global independent move})
{
\begin{enumerate}
\item[7a.] Sample proposal $\tilde{\bm{\beta}}^* \sim N\left(\tilde{\bm{\beta}};\tilde{\bm{\beta}}^{ML},\Sigma^{ML}\right)$.\;
\item[7b.] Construct proposal $\bm{\beta}^* = [I_{r \times r},\tilde{\bm{\beta}}^*]$.\;
\end{enumerate}
}
\begin{enumerate}
\item[8.]{Calculate Metropolis Hastings Acceptance Probability:
\begin{equation}
A\left(\bm{\beta}^{(j-1)},\bm{\beta}^*\right) = \frac{p\left(\Sigma^{(j)}, B^{(j)}, \bm{\beta}^*|Y\right)q\left(\bm{\beta}^* \rightarrow \bm{\beta}^{(t-1)}\right) }
{p\left(\Sigma^{(j)}, B^{(j)}, \bm{\beta}^{(j-1)}|Y\right)q\left(\bm{\beta}^{(t-1)} \rightarrow \bm{\beta}^*\right)}
\end{equation}
Accept $\bm{\beta}^{(j)}=\bm{\beta}^*$ via rejection using A, otherwise $\bm{\beta}^{(j)}=\bm{\beta}^{(j-1)}$.}\;
\item[9.] $j = j + 1$\;
\end{enumerate}
}
}
\caption{MH within Gibbs sampler for fixed rank r via a pretuned mixture of global and local moves.\label{IR}}
\end{algorithm}
}}
\subsection{Algorithm 2: Adaptive Metropolis within Gibbs sampler moves for CVAR model given rank r}

There are several classes of adaptive MCMC algorithms, see Roberts
and Rosenthal (2009). The distinguishing feature of adaptive MCMC
algorithms, compared to standard MCMC, is generation of the Markov
chain via a sequence of transition kernels. Adaptive algorithms
utilize a combination of time or state inhomogeneous proposal
kernels. Each proposal in the sequence is allowed to depend on the
past history of the Markov chain generated, resulting in many
variants.

Due to the inhomogeneity of the Markov kernel used in adaptive
algorithms, it is particularly important to ensure the generated
Markov chain is ergodic, with the appropriate stationary
distribution. Several recent papers proposing theoretical
conditions that must be satisfied to ensure ergodicity of adaptive
algorithms include, Atachade and Rosenthal (2005), Roberts and
Rosenthal (2009), Haario {\textit{et al.}} (2007), Andrieu and Moulines
(2006) and Andrieu and Atachade (2007).

Haario {\textit{et al.}} (2001) developed an adaptive Metropolis algorithm with proposal covariance adapted to the history of the Markov chain. The original proof of ergodicity of the Markov chain under such an adaption was overly restrictive. It required a bounded state space and a uniformly ergodic Markov chain.

Roberts and Rosenthal (2009) proved ergodicity of adaptive MCMC under simpler conditions known as \textit{Diminishing Adaptation} and \textit{Bounded Convergence}. As in Roberts and Rosenthal (2009) we assume that each fixed kernel in the sequence $Q_{\gamma}$ has stationary distribution $P\left(\cdot\right)$. Define the convergence time for kernel $Q_{\gamma}$ when starting from state $\theta$ as $M_{\epsilon}\left(\theta,\gamma\right) = \text{inf}\{j \geq 1 : \|Q^j_{\gamma}\left(\theta,\cdot\right) - P\left(\cdot\right)\| \leq \epsilon \}$.
Under these assumptions, they derive the sufficient conditions;
\begin{itemize}
\item{ \textbf{Diminishing Adaptation:} $\text{lim}_{n\to\infty}\text{sup}_{\theta \in E}\|Q_{\Gamma_{j+1}}\left(\theta,\cdot\right) - Q_{\Gamma_{j}}\left(\theta,\cdot\right)\| = 0$ in probability. Note, $\Gamma_j$ are random indices.}
\item{ \textbf{Bounded Convergence:} $\{M_{\epsilon}\left(\Theta^{(j)},\Gamma_j\right)\}^\infty_{j=0}$ is bounded in probability, $\epsilon > 0.$}
\end{itemize}
which guarantee asymptotic convergence in two senses,
\begin{itemize}
\item{Asymptotic convergence: $\text{lim}_{j\to\infty}\|\mathcal{L}\left(\Theta^{(j)}\right)-P\left(\cdot\right)\|=0$}
\item{WLLN: $\text{lim}_{j\to\infty}\frac{1}{j}\sum^{j}_{i=1}g\left(\Theta^{(i)}\right)=\int g(\theta)p(\theta)d\theta$ for all bounded $g : E \to \R$.}
\end{itemize}

It is non-trivial to develop adaption schemes which can be verified to satisfy these two conditions. We develop a matrix-variate adaptive MCMC methodology in the CVAR setting, using a proposal kernel known to satisfy these two ergodicity conditions for unbounded state spaces and general classes of target posterior distribution, see Roberts and Rosenthal (2009) for details.

In Algorithm 2 the mixture proposal distribution for parameters
$\tilde{\bm{\beta}}$ which is $d=\left(n-r\right) \times r$ dimensional
and is given at iteration $j$ by,
\begin{equation}
q_j\left(\tilde{\bm{\beta}}^{(t-1)},\cdot\right)= w_1 N\left(\tilde{\bm{\beta}};\tilde{\bm{\beta}}^{(t-1)},\frac{\left(2.38\right)^2}{d}\Sigma_j\right) + \left(1-w_1\right) N\left(\tilde{\bm{\beta}};\tilde{\bm{\beta}}^{(t-1)},\frac{\left(0.1\right)^2}{d}I_{d,d}\right).
\end{equation}
Here, $\Sigma_j$ is the current empirical estimate of the
covariance between the parameters of $\tilde{\bm{\beta}}$ estimated
using samples from the Markov chain up to time $j$. The
theoretical motivation for the choices of scale factors 2.38, 0.1
and dimension d are all provided in Roberts and Rosenthal (2009)
and are based on optimality conditions presented in Roberts {\textit{et al.}}
(1997) and Roberts and Rosenthal (2001). The adaptive MCMC
Algorithm 2 is identical to Algorithm 1 except we replace step 7
with the following alternative;\newline
\rule{\textwidth}{1pt}
\\
\textbf{Algorithm 2}: matrix-variate adaptive MH within Gibbs sampler for fixed rank r.\label{IR}\\
\rule{\textwidth}{1pt}
\\
\eIf(\tcc*[f]{perform an adaptive random walk move}){$u \geq w_1$}{
\begin{enumerate}
\item[7a.] Estimate $\Sigma_j$ the empirical covariance of $\bm{\beta}$ for elements in $(n-r) \times r$ using samples $\{\tilde{\bm{\beta}}^{(i)}\}_{i=1:j}$.\;
\item[7b.] Sample proposal $\tilde{\bm{\beta}}^* \sim N\left(\tilde{\bm{\beta}};\tilde{\bm{\beta}}^{(t-1)},\frac{\left(2.38\right)^2}{d}\Sigma_j\right)$.\;
\item[7c.] Construct proposal $\bm{\beta}^* = [I_{r \times r},\tilde{\bm{\beta}}^*]$.\;
\end{enumerate}
}(\tcc*[f]{perform a non-adaptive random walk move})
{
\begin{enumerate}
\item[7a.] Sample proposal $\tilde{\bm{\beta}}^* \sim N\left(\tilde{\bm{\beta}};\tilde{\bm{\beta}}^{(t-1)},\frac{\left(0.1\right)^2}{d}I_{d,d}\right)$.\;
\item[7b.] Construct proposal $\bm{\beta}^* = [I_{r \times r},\tilde{\bm{\beta}}^*]$.\;
\end{enumerate}
}
\rule{\textwidth}{1pt}

\section{Rank Estimation for Bayesian VAR Cointegration models}
\label{modelCVAR4}

Here we discuss the Bayes Factor approach to rank estimation, noting that it is computationally inefficient, since it involves running n+1 Markov chains, one for each model (rank $r$). For a sophisticated alternative which presents a novel TD-MCMC based approach, requiring a single Markov chain to obtain samples from the posterior distribution $p\left(B,\Sigma,\bm{\beta},r|Y\right)$, see Peters \textit{et al.} (2009).
\subsection{Posterior Model Probabilities for Rank $r$ via Bayes Factors}

In Sugita (2002) and Kleibergen and Paap (2002) the rank is
estimated via Bayes factors, a popular approach to Bayesian model
selection in Bayesian cointegration literature. We note that
alternative approaches to rank estimation include Strachan and van
Dijk (2004) and Strachan and van Dijk (2007). Sugita (2002) works with a conjugate prior on
$\bm{\alpha}$ which will not produce a problem with Bartlett's paradox,
posterior probabilities of the rank are well defined.
\subsubsection{Bayes Factors}

The earlier work of Sugita (2002) compares the rank of the unrestricted $\bm{\alpha}$ to the 0 rank setting. Note, Kleibergen and Pap (2002) have a slightly different approach in that they compared each rank $r$ to the full rank case for the unrestricted $\bm{\alpha}$ parameter. Recently, Sugita (2009) revisits the important question of rank estimation via Bayes Factors also comparing the Schwarz BIC approximation and Chib's (1995) approach for the marginal likelihood. 

Under a rank 0 comparison, the posterior model probabilities are given by,
\begin{equation}
Pr\left(r|Y\right) = \frac{BF_{r|0}}{\sum^n_{j=0}BF_{j|0}},
\end{equation}
with $BF_{0|0}$ defined as 1.

In the calculation of $BF_{r|0}$, Sugita (2002) recommends an approach first introduced by
Verdinelli and Wasserman (1995) for nested model structure Bayes
factors, which results in

\begin{equation}
BF_{r|0} = \frac{p(\bm{\alpha}'=\bm{0}_{r \times n})}{C_{r}^{-1} p(\bm{\alpha}'=\bm{0}_{r \times n}|Y)} =\frac{\int p(\bm{\alpha},\bm{\beta},\Gamma,\Sigma|Y) d\bm{\alpha} d\bm{\beta} d\Gamma d\Sigma}{C_{r}^{-1} \dot \int p(\bm{\alpha},\bm{\beta},\Gamma,\Sigma|Y)|_{rank(\bm{\alpha})=0} d\bm{\alpha} d\bm{\beta} d\Gamma d\Sigma}
\end{equation}
where the correction factor for the reduction in dimension $C_{r}$ is given by,
\begin{equation}
C_{r} = \int p(\bm{\alpha},\bm{\beta},\Gamma,\Sigma)|_{rank(\bm{\alpha})=0} d\bm{\beta} d\Gamma d\Sigma.
\end{equation}

We note that Sugita (2002) does not comment on numerical
complications that can arise when implementing this
estimator for the CVAR model. We detail in Appendix 1, Section
\ref{BFNumerics} steps that were critical to the calculation of
the Bayes Factors when handling potential numerical overflows. The
numerical issues arise as $t$ increases, for example the term
$|S_{*}^{(i)}|^{\frac{t+h}{2}}$ will explode numerically. This
will result in incorrect numerical results for the Bayes Factors
if not handled appropriately.

\subsection{Model Selection, Model Averaging and Prediction}

With samples from $p(\bm{\beta}, B,\Sigma,r|Y)$ one can consider either
model selection or model averaging. In a survey of the literature
on rank selection, the most common form of inference performed
involves model selection. In this paper we note that model
averaging should also be considered, especially when it is
probable that given the realized data, two different ranks are
highly probable according to their posterior model probabilities. We argue that by adopting 
the Bayesian model averaging framework one is able to reduce potential model risk associated with 
selection of the rank from several choices, which may all be fairly probable under the posterior. 
This in turn should reduce the associate model risk involved in the popular application of CVAR 
models in algorithmic trading strategies based on these co-integration frameworks and estimation of the rank.

In this case one can use the samples from $p(\bm{\beta}, B,\Sigma|Y,r)$
in each model $r$ to form a weighted model averaged estimate
through the direct knowledge of the estimated model probabilities
given by $p(r|Y)$. There is discussion on model averaging in the
CVAR context found in Koop \textit{et al.} (2006).\newline 

\noindent \textbf{Bayesian Model Order Selection (BMOS)}
\newline In BMOS we select the most probable model corresponding to
the maximum \textit{a posteriori} (MAP) estimate from $p(r|Y)$,
denoted $r_{MAP}$. Conditional on $r_{MAP}$, we then take the
samples of $\{\bm{\beta}^{(j)}, B^{(j)}, \Sigma^{(j)}\}_{j=1:M}$
corresponding to Markov chain simulated for the $r_{MAP}$ model and we estimate point estimates for the
parameters.

These point estimates typically include posterior means or modes,
though one should be careful. We note that it was demonstrated by
Kleibergen and van Dijk (1994) or Bauwens and Lubrano (1996) that
in many popular CVAR Bayesian models, certain choices of prior
result in a proper posterior yet it may not have finite moments of
any order. Some alternatives are proposed by Strachan and Inder
(2004). \newline
\\
\textbf{Bayesian Model Averaging (BMA)} \newline
In this section we consider the problem of estimating for example
an integral of a quantity or function of interest, $\phi(\{\bm{\beta},
B, \Sigma\})$, with respect to the posterior distribution of the
parameters, e.g. moments of the posterior. Since we have chosen to
work with a posterior distribution $p(\bm{\beta}, B, \Sigma, r|Y)$ we
can estimate this integral quantity whilst removing the model risk
associated with rank uncertainty. This is achieved by
approximating
\begin{equation}
\begin{split}
\sum_{r=1}^{n}\int \phi(\{\bm{\beta}, B, \Sigma|r\}) p(\bm{\beta}, B, \Sigma|Y,r)p(r|Y) d\bm{\beta} dB d\Sigma \approx \sum_{r=1}^{n}\sum_{j=1}^{M} \phi(\{\bm{\beta}^j, B^j, \Sigma^j|r^j\})p(r^j|Y).
\end{split}
\end{equation}
\textbf{Prediction Incorporating Model Risk} \newline
Here we perform prediction whilst removing model uncertainty related to the rank. This is possible under a Bayesian Model Averaging (BMA) framework using,
\begin{equation}
p(Y^*|Y) = \sum_{r=1}^{n} \int p(Y^*|\bm{\beta}, B, \Sigma, r)p(\bm{\beta}, B, \Sigma|Y,r)p(r|Y) d\bm{\beta} dB d\Sigma.
\end{equation}
We will compare the predictive performance of the MMSE estimate or mean of the estimated distribution for $p(Y^*|Y)$ under the BMA versus BMOS approach which involves,
\begin{equation}
p(Y^*|Y) = \int p(Y^*|\bm{\beta}, B, \Sigma, \widehat{r}^{\text{MAP}})p(\bm{\beta}, B, \Sigma|Y,\widehat{r}^{\text{MAP}}) d\bm{\beta} dB d\Sigma.
\end{equation}

\section{Simulation Experiments}
Analysis of the methodology developed is in three parts: the first part contains simulations performed on synthetic data sets, comparing performance of the proposed model sampling methodology; the second part contains two real data set examples; and the third part involves analysis of predictive performance BMOS and BMA using real data.

\subsection{Synthetic Experiments}
In this section the intention will be to develop a controlled setting in which the true model parameters are known and the data is generated from the true model. This will allow us to assess performance of each of the proposed estimation procedures. In doing this we take an identical model to the simple model studied in Sugita (2002; 2009) [p.4] for our analysis.

\subsubsection{Analysis of samplers}

The first analysis is to compare the performance of the two
adaptive samplers. To achieve this we generate 20 realizations of
data sets of length $T=100$ from the rank $r=2$ model. Then
conditional on knowledge of the rank $r=2$ we sample $J=20,000$
samples from the joint posterior
$p\left(B,\Sigma,\bm{\beta}|Y,r=2\right)$ and discard the first 10,000
samples as burnin. We perform this analysis for each of the data
realizations under both of the proposed samplers, Algorithm 1 and
Algorithm 2, and then we present average MMSE estimates and
average posterior standard deviations from each sampler in Table
1. In particular we present the averaged posterior point estimates
for: the unrestricted $\bm{\beta}$ parameters; the average trace of the
posterior estimate of the covariance $\Sigma$; the average of each
of the intercept terms; and the averaged first element of the
unrestricted $\bm{\alpha}$.

Note, the pre-tuning of the local random walk proposal standard
deviation for Algorithm 1 is performed offline using an MCMC run
of length 20,000. Additionally, the prior parameters were set to
be: for $B|\Sigma$ the prior parameters were set as $P =
\left(\widehat{W}'\widehat{W}\right)^{-1}\widehat{W}Y$,
$A=\lambda\left(\widehat{W}'\widehat{W}\right)/T$ with $\lambda
=1$, $\widehat{W}=\left(X Z\widehat{\bm{\beta}}\right)$ and
$\widehat{\bm{\beta}} = [I_{r},\mathbf{0}]$;  for $\bm{\beta}$ the
prior parameters were set as $E[\bm{\beta}] = \left(I_{r},\mathbf{0}\right)$, $Q = I_{n}$, $H=\tau Z'Z$ and
$\tau=1/T$; for $\Sigma$ the prior parameters were set as $S=\tau
Y'Y$ and $h=n+1$.

These results demonstrate that both Algorithm 1 and Algorithm 2
perform well. The MMSE estimates produced by both algorithms are
accurate compared to the true parameter values used to generate
the data. Algorithm 1 which involved the mixture of pretuned local
moves and a Global move centered on the Maximum Likelihood
parameter estimates required more computational effort than the
adaptive MCMC approach of Algorithm 2. Additionally, we point out
that as discussed in Rosenthal (2008), the sampler we developed in
Algorithm 2 actually achieves optimal performance as $n
\rightarrow \infty$. Therefore it will be a far superior algorithm
to the griddy Gibbs sampler approach which will not be feasible in
high dimensions. Hence, for an automated and computationally
efficient alternative to the griddy Gibbs sampler typically used
we would recommend the use of Algorithm 2. In the following
studies, we utilize Algorithm 2, the adaptive MCMC algorithm. To conclude, 
we also present the trace plots of the sample paths under the adaptive MCMC algorithm, 
see Figure 1. This plot demonstrates that rapid convergence of the MMSE estimates 
of the parameters in the posterior, even when initialized far from the true values. Additionally,
one can see the behavior of the adaptive proposal, learning the appropriate proposal variance.

\subsubsection{Analysis of Adaptive MCMC sampler in high dimension.}
In this example, we work with the Adaptive MCMC algorithm we developed for the Bayesian CVAR model.
In particular we consider the case in which $n = 10$, which is a setting in which the standard approach of 
the griddy Gibbs sampler will become excessively computational, due to the curse of dimensionality, since there are now several hundred parameters to be 
sampled from the posterior. 

All coefficients except for the cointegrating vectors are generated by uniform distributions with a range between -0.4 to 0.4, and the error
covariance was set to the identity. We generate a realizations of data of length $T=100$ from the true rank $r=5$ model in which the cointegration vector 
has all terms in the matrix of $\bm{\beta}$ which are unrestricted set to be 0, other than the last row, which is -1. 
Then conditional on knowledge of the rank $r=5$ we sample $J=20,000$ samples from the joint posterior 
$p\left(B,\Sigma,\bm{\beta}|Y,r=5\right)$ and discard the first 10,000 samples as burnin. 

The sample paths of the cointegration vector parameters randomly selected to be presented were $\beta_{10,1},\beta_{10,4}$ which are shown in Figure 2.
Clearly, again in this high dimensional setting (310 dimensions), the adaptive MCMC algorithm performs suitably. Even, though the Markov chain is initialized far from 
the true parameter values of cointegration vector, we see the rapid convergence of our sampler. This is illustrated for the two arbitrarily selected parameters
which had true values of of -1 and -1. Note, in this high dimensional setting, the algorithm was implemented in Matlab and took only 132sec to complete the simulation on an Intel Core 2 Duo at 2.40GHz, with 3.56Gb of RAM.

\subsubsection{Analysis of model selection in the Bayesian CVAR model}
In this section we study on synthetic data the performance of
the Bayes Factor estimator applied to estimate posterior
model probabilities for the rank. To perform this analysis we consider the model from Sugita, (2002; 2009 [p.4]) and we take data series of
length $T = 100$ and we simulate 50 independent data realizations
for each possible model rank $r=1,\ldots,4$. Then for each rank
$r$ we count the number of times each model is selected as the MAP
estimate out of the total of the 50 simulations, one simulation
per generated data set. Note, the algorithm was run for 20,000
iterations with 10,000 samples used as burnin. The results of this
analysis are presented in Table 2.

We note that the results of this section demonstrated the
following interesting properties:
\begin{enumerate}
\item{ When the true rank used to generate the observations data was small, the BF methodology was clearly able to detect the true model order as the MAP estimate in a high proportion of the tested data sets.}
\item{ In all cases the averaged actual posterior model probabilities were very selective of the correct model, indicating that at least under this synthetic data scenario, there would not be great benefit in performing model averaging. However, we will demonstrate later examples with actual data in which there is significant ambiguity between possible model ranks, in these cases we also study the model averaging results.}
\end{enumerate}

\subsection{Financial Example 1 - US mini indexes}
\label{FinancialExample1 - } Having assessed the proposed
algorithms developed in this paper for synthetic data generated
from a CVAR model, we now work with a practical financial example.
In this example we will consider data series comprised of US
indexes S\&P mini, Nasdaq mini and Dow Jones mini. The data
obtained for each of these data series consists of 774 values
corresponding to the close of market daily price from the
31-Aug-2005 through to 30-Sep-2008. The time series data is
presented in Figure 3. 

We analyze this data using Algorithm 2 (adaptive MCMC) and
estimate the rank via Bayes Factor analysis, the results are
presented in Table 3. We run 20 independent samplers with different
initializations, for each possible rank. This is performed for
each data set, and the total series is split into increasing
subsets, each taking subsets of the data from 50 data points
through to 400 data points, in increases of 50 data points. This
allows us to study the change in the estimated rank as a function
of time for each of these time series. Clearly, if the true rank
of our model was fixed, then as the total amount of data we
include increases, then we should see the posterior model
probability of the rank converge to 1 for one of the possible
ranks. What we observed after doing this analysis was that there
was a clear variability in the predicted rank as we included more
data. In particular the model estimates showed preference most often to rank
1, suggesting that 2 common stochastic trends are present in the series.
Additionally, the fact that in several cases, the model is less
likely to distinguish between rank 1 and 2, suggests it may be
prudent to also perform a model averaging analysis. Especially in the popular 
application of CVAR models in practice to perform algorithmic trading.

\subsection{Financial Example 2 - US notes}

Here we repeat the same procedure performed in Financial Example 1, for a different data set. This time we consider data series comprised of Bond data for US 5 year, 10 year and 30 year notes over the same time period as the US mini index data. The time series data is presented in Figure 4. 

We analyze this data using Algorithm 2 (adaptive MCMC) and
estimate the rank via Bayes Factor analysis, the results are
presented in Table 4. We set up this second data analysis in the same way as
Financial Exampl 1, with 20 independent samplers, each with different
initializations, for each possible rank. This allows us to study the 
change in the estimated rank as a function
of time for each of these time series. Again, we observed that
with this data, the model gave preference most often to rank 1,
suggesting that 2 common trends are present in the series we are
analyzing. However, there was much stronger evidence for a single
co-integrating relationship over time in this data, compared to
the analysis of the US mini index data over the same period. This
suggests that the US bond data series is a more stable series to
fit the CVAR model too when assuming a constant number of
co-integrating relationships over time.

\subsection{Financial Example 3}

In this section we perform a predictive performance comparison
using Bayesian Model Selection versus Averaging. We take 2 series
for the US bonds, 5 years and 10 years, and we combine these
series over the same period with the S\& P 500 mini index. We
compare the MMSE estimate of the predicted series over 10 steps
ahead which is obtained from the distribution of the predicted
data $p\left(Y^*|Y\right)$, after we have integrated out parameter
and rank uncertainties. We demonstrate that in this actual data
example, the performance obtained by Bayesian Model Averaging
represents the uncertainty in the prediction more accurately than
the Bayesian Model Order Selection setting.

This study is performed as follows. We begin by selecting
randomly, with replacement, 100 segments of the vector time
series, each containing 50 days of data. For each segment of the
time series we fit our Bayesian model for each possible rank, also
estimating via Bayes Factors the posterior model probability for
each rank. Then we calculate the predictive posterior mean,
corresponding to the MMSE estimate of the predicted data series
for the following 5 days, $Y^*$. Finally, we take the squared
difference between the actual data series over the proceeding 10
days post the 50 days for the given segment and the posterior mean
of the predicted data $Y^*$.

In Figure 5 we present for each prediction day a boxplot of the squared difference between the actual data over the random sets of 5 days and the predictive posterior MMSE estimators for the same 5 days. We compare here the performance under Bayesian model selection and averaging. When performing Bayesian model averaging we are integrating out uncertainty in the prediction due to the prediction of the unknown rank.

Clearly, the Bayesian model averaging approach will result in a greater uncertainty in the prediction when compared to the Bayesian model selection. This is reflected especially in the distribution of the prediction at 5 days where the model averaging approach box-whisker plot covers a noticeably wider range than the model selection equivalent. Though not presented here, we also assessed and confirmed this would occur out to longer predictions of 10 days and 20 days.

\section{Conclusions}

We have developed and demonstrated how one can utilize state of the art adaptive MCMC methodology to solve a challenging high dimensional econometrics problem based on cointegrated vector autoregressions. The challenging application involved a posterior distribution which was matrix-variate and very high dimensional. We compared the performance of the Adaptive Metropolis algorithm with an alternative based on a mixture proposal of local and global moves centered on the the Maximum Likelihood parameters. We then formulated the rank estimation in as a Bayesian model selection problem and performed analysis of the Bayes factors using our adaptive MCMC algorithm. We concluded with analysis of real market data and performed Bayesian model selection and model averaging, with respect to the unknown rank. In conclusion, the adaptive MCMC methodology developed clearly allowed us to extend significantly the dimension of the estimation problem in the Bayesian CVAR literature. It was shown to be highly efficient and accurate.

From the perspective of developing a Bayesian CVAR model for algorithmic trading we found that historically the US bond data we considered is a more stable series to fit the CVAR model too when assuming a constant number of co-integrating relationships over time. This will therefore impact the stability of trading performance under such models. In addition when considering trading triples made up of the US bond data series and the S\&P mini index, it is beneficial to perform Bayesian model averaging for the rank, rather than just selecting the most probable co-integration rank. The adaptive MCMC based framework allows this to be done efficiently and in an automated fashion.

\section{Acknowledgments}
We would like to thank Prof. Robert Kohn for useful feedback and also we thank the two anonymous referees and associate editor for their very helpful comments which have significantly improved the exposition of the paper. The first author thanks the Statistics Department of the University of NSW for financial support and Boronia Managed Funds.  

\section{References}
{\small{
\begin{enumerate}


\item{Andrieu, C., Moulines, E. (2006). "On the ergodicity properties of some adaptive MCMC algorithms". \textit{Annals of Applied Probability}, 16, (3).}

\item{Andrieu, C., Atachade, Y. (2007). "On the efficiency of adaptive MCMC algorithms". \textit{Electronic Communications in Probability}, 12, 336-349.}

\item{Atachade, Y.F., Rosenthal, J.S. (2005). "On adaptive Markov chain Monte Carlo algorithms". \textit{Bernoulli}, 11(5), 815-828.}


\item{Bauwens, L., Lubrano, M. (1996). "Identification restrictions and posterior densities in cointegrated Gaussian VAR systems". \textit{Advances in Econometrics} 11, Part B, (JAI Press, Greenwich) 3-28.}

\item{Bauwens, L., Giot, P. (1997). "A Gibbs sampling approach to cointegration". \textit{Computational Statistics}, 13, 339-368.}

\item{Chib, S, (1995). "Marginal likelihood from the Gibbs output". \textit{Journal of the American Statistical Association}, 90(432), Theory and Methods, 1313-1321.}

\item{Engle, R.F., Granger, C.W.J. (1987). "Co-Integration and Error Correction: Representation, Estimation, and Testing". \textit{Econometrica}, 55(2), 251-276.}

\item{Fan, Y., Peters, G.W., Sisson, S.A. (2009). "Automating and evaluating reversible jump MCMC proposal distributions". \textit{Statistics and Computing}, 19, 409-421.}

\item{Geweke J. (1996). "Bayesian reduced rank regression in econometrics". \textit{Journal of Econometrics}, 75, 121-146.}

\item{Giordani, P. and Kohn, R. (2006). "Adaptive independent Metropolis-Hastings by fast estimation of mixtures of normals." Preprint.}

\item{Granger, C.W.J. (1981). "Some properties of time series data and their use in econometric model specification". \textit{Journal of Econometrics}, 121-130.}

\item{Granger, C.W.J., Weiss, A.A. (1983). "Time series analysis of error-correcting models". \textit{Studies in econometrics, time series, and multivariate statistics.} New York: Academic Press, 255-278.}

\item{Haario, H., Saksman, E., Tamminen, J. (2007). "Componentwise adaptation for high dimensional MCMC". \textit{Computational Statistics}, 20(2). }

\item{Haario, H., Saksman, E., Tamminen, J. (2001). "An adaptive metropolis algorithm". \textit{Bernoulli}, 7, 223-242. }

\item{Kleibergen, F., Paap, R. (2002). "Priors, posteriors and Bayes factors for a Bayesian analysis of cointegration". \textit{Journal of Econometrics}, Elsevier, 111(2), 223-249.}

\item{Kleibergen, F., van Dijk, H.K. (1994). "On the Shape of the Likelihood/Posterior in Cointegration Models". \textit{Econometric Theory}, Cambridge University Press, 10(3-4), 514-551.}

\item{Koop, G., Strachan, R., van Dijk, H., Villani, M. (2006). "Bayesian Approaches to Cointegration". In T.C.Mills and K. Patterson (Ed.), \textit{Palgrave Handbook of Econometrics Volume 1 Econometric Theory 1st ed.} (pp. 871-898) UK: Palgrave Macmillan.}


\item{L\"{u}tkepohl, H. (2007). \textit{New Introduction to Multiple Time Series Analysis}. Springer}

\item{Peters, G.W., Kannan, B.K., Lasscock, B., Mellen, C. (2009). "Trans-dimensional and adaptive Markov chain Monte Carlo for Bayesian co-integrated VAR models". Technical Report: University of NSW, Statistics Department.}

\item{Reinsel, G.C., Velu, R.P. (1998). \textit{Multivariate Reduced-Rank Regression, Theory and Applications}. Lectuer Notes in Statistics Springer.}

\item{Roberts, G.O., Gelman, A., Gilks, W.R. (1997). "Weak convergence and optimal scaling of random walk Metropolis algorithms". \textit{Annals of Applied Probability}, 7, 110-120.}

\item{Roberts, G.O., Rosenthal, J.S. (2009). "Examples of Adaptive MCMC". \textit{Journal of Computational and Graphical Statistics}, 18(2), p.349-367. }

\item{Roberts, G.O., Rosenthal, J.S. (2001). "Optimal scaling for various Metropolis-Hastings algorithms". \textit{Statistical Science}, 16, 351-367.}

\item{Rosenthal, J.S. (2008). "Optimal Proposal Distributions and Adaptive MCMC". {\textit Chapter for MCMC Handbook}, Brooks S., Gelman A., Jones G. and Meng X.L. eds.}

\item{Silva, R., Giordani, P., Kohn, R., Pitt, M. (2009). "Particle filtering within adaptive Metropolis Hastings sampling." Arxiv preprint arXiv:0911.0230.}

\item{Strachan, R.W., van Dijk, H.K. (2004). "Bayesian model selection with an uninformative prior". \textit{Keele Economics Research
Papers}}

\item{Strachan, R.W., van Dijk, H.K. (2007). "Bayesian model averaging in Vector Autoregressive Processes with an Investigation of Stability of the US Great Ratios and Risk of a Liquidity Trap in the USA, UK and Japan".}

\item{Strachan, R.W., Inder, B. (2004). "Bayesian analysis of the error correction model". \textit{Journal of Econometrics}, 123(2), 307-325.}

\item{Sugita, K. (2002). "Testing for cointegration rank using Bayes factors". \textit{Royal Economic Society Annual Conference}, Royal Economic
Society.}

\item{Sugita, K. (2009). "A Monte Carlo comparison of Bayesian testing for cointegration rank". \textit{Economics Bulletin}, 29(3), 2145-2151.}

\item{Verdinelli, I. and Wasserman, L. (1995). "Computing Bayes Factors Using a Generalization of the Savage-Dickey Density Ratio". \textit{Journal of the American Statistical Association}, 90(433).}

\item{Vermaak,J., Andrieu, C., Doucet, A., Godsill, S.J. (2004). "Reversible jump Markov chain Monte Carlo strategies for Bayesian model selection in autoregressive processes". \textit{Journal of Time Series Analysis}, 25(6), 785-809.} 


\end{enumerate}
}}
\newpage
\section{Appendix 1}
\label{BFNumerics}

We begin by calculating the log posterior model probabilities,
\begin{equation}
\text{log}\left(Pr\left(r|Y\right)\right) = \text{log}\left( BF_{r|0} \right) + \text{log}\left(BF_{max|0}\right) - \text{log}\left(\sum^n_{j=0}\text{exp}\left( \text{log}\left(BF_{j|0} \right) - \text{log}\left(BF_{max|0} \right) \right) \right),
\end{equation}
where $BF_{max|0} = max\{BF_{0|0},...,BF_{n|0}\}$. Additionally, we now consider the log of the Bayes Factor for rank $r$ and we apply the same numerical trick.
\begin{equation}
\text{log}\left(BF_{r|0}\right) = \text{log}\left(p(\bm{\alpha}'=0_{r \times n})\right) + \text{log}\left(C_r\right) - \text{log}\left(p(\bm{\alpha}'=0_{r \times n}|Y)\right)
\end{equation}

Now, considering each of the terms:
\begin{itemize}

\item $\text{log}\left(p(\bm{\alpha}'=0_{r \times n})\right)=-\frac{nr}{2}\text{log}(\pi) + \frac{h}{2}\text{log}\left(|S|\right) + \frac{n}{2}\text{log}\left(|A_{22.1}|\right) +\sum_{j=1}^{n}\text{log}\left(\frac{\Gamma(\frac{h+r+1-j}{2})}{\Gamma(\frac{h+1-j}{2})}\right) -\frac{h+r}{2}\text{log}\left(|S|\right)$

\item $\text{log}\left(p(\bm{\alpha}'=0_{r \times n}|Y)\right)=
-\text{log}\left(N\right) + \text{log}\left(L^{(1)}_{max}\right)
-\text{log}\left( \text{exp}\left( \sum^N_{i=1} \text{log}\left(L^{(1)}_i\right) - \text{log}\left(L^{(1)}_{max}\right) \right) \right)$, \newline
where $L^{(1)}_i = \pi^{-\frac{nr}{2}} |S_{*}^{(i)}|^{\frac{t+h}{2}} |A_{*22,1}^{(i)}|^{\frac{n}{2}} {\prod_{i=1}^{n}\frac{\Gamma(\frac{t+h+r+1-i}{2})}{\Gamma(\frac{t+h+1-i}{2})}}|S_{*}^{(i)} + B_{*2}^{(i)'} A_{*22.1}^{(i)} B_{*2}^{(i)}|^{- \frac{t+r}{2}}$ and $L^{(1)}_{max} = \text{max}\{L^{(1)}_1,...,L^{(1)}_N\}$.

\item $\text{log}\left(C_{r}\right) = -\text{log}(N) + \text{log}\left(L^{(2)}_{max}\right) \text{log}\left( \text{exp}\left( \sum^N_{i=1} \text{log}\left(L^{(2)}_i\right) - \text{log}\left(L^{(2)}_{max}\right) \right) \right)$, \newline
where $L^{(2)}_i =\frac{p(\bm{\alpha}=0,\Gamma^{(i)}|\Sigma^{(i)})}{p(\Gamma^{(i)}|\Sigma^{(i)})}$ and $L^{(2)}_{max} = \text{max}\{L^{(2)}_1,...,L^{(2)}_N\}$.
Note this sum evaluated using samples from the Markov chain run in model $r$ where, $p(\bm{\alpha}=0,\Gamma^{(i)}|\Sigma^{(i)})$ and $p(\Gamma^{(i)}|\Sigma^{(i)})$ are obtained using knowledge of the specified prior, $p(B|\Sigma)=p(\Gamma,\bm{\alpha}|\Sigma)=p(\mu,\Psi_{1:p-1},\bm{\alpha}|\Sigma)$.
\end{itemize}

\newpage
\begin{table}
{\centering {\small{
\begin{tabular}{|c|c|c|c|}
    \hline
    \textbf{Parameter Estimates}   & \textbf{Algorithm 1} & \textbf{Algorithm 2} & \textbf{Truth} \\ \hline
    Ave. MMSE $\beta_{1,r+1}$           & -0.002 (0.001) & -0.034 (0.002)   & 0\\ \hline
   Ave. Posterior Stdev. $\beta_{1,r+1}$ & 0.018 (0.006) & 0.010 (0.003)    & -\\ \hline
    Ave. MMSE $\beta_{2,r+1}$           & -0.819 (0.051) & -0.862 (0.045)   & -1\\ \hline
   Ave. Posterior Stdev. $\beta_{2,r+1}$ & 0.032 (0.005) & 0.020 (0.003)    & - \\ \hline
    Ave. MMSE $\beta_{1,n}$             & 0.033 (0.025) & -0.024 (0.023)    & 0\\ \hline
   Ave. Posterior Stdev. $\beta_{1,n}$  & 0.030 (0.012) & 0.026 (0.010)     & -\\ \hline
    Ave. MMSE $\beta_{2,n}$             & -0.752 (0.098) & -0.774 (0.082)   & -1\\ \hline
   Ave. Posterior Stdev. $\beta_{2,n}$
                                                & 0.038 (0.013) & 0.028 (0.006)     & -\\ \hline
   Ave. Mean acceptance probability $\bm{\beta}$
                                                & 0.352 (0.010) & 0.232 (0.029)     & -\\ \hline\hline
    Ave. MMSE $\text{tr}\left(\Sigma\right)$
                                                & 4.945 (0.331) & 4.432 (0.332)     & 4\\ \hline
   Ave. Posterior Stdev. $\text{tr}\left(\Sigma\right)$
                                                & 0.420 (0.049) & 0.416 (0.048)     & - \\ \hline\hline
    Ave. MMSE $\mu_1$                   & 0.07 (0.051) & 0.065 (0.043)  & 0.1\\ \hline
   Ave. Posterior Stdev. $\mu_1$            & 0.236 (0.028) & 0.226 (0.026)     & -\\ \hline
    Ave. MMSE $\mu_2$                   & -0.027 (0.041) & -0.034 (0.024)   & 0.1\\ \hline
   Ave. Posterior Stdev. $\mu_2$            & 0.183 (0.041) & 0.181 (0.010)     & - \\ \hline
    Ave. MMSE $\mu_3$                   & -0.080 (0.084) & -0.061 (0.045)   & 0.1\\ \hline
   Ave. Posterior Stdev. $\mu_3$            & 0.199 (0.020) & 0.187 (0.015)     & -\\ \hline
    Ave. MMSE $\mu_4$                   & 0.024 (0.049) & 0.030 (0.029)     & 0.1\\ \hline
   Ave. Posterior Stdev. $\mu_4$            & 0.184 (0.010) & 0.185 (0.011)     & -\\ \hline\hline
    Ave. MMSE $\alpha_{1,1}$            & -0.223 (0.015) & -0.224 (0.016)   & -0.2 \\ \hline
   Ave. Posterior Stdev. $\alpha_{1,1}$     & 0.070 (0.006) & 0.068 (0.005)     & -\\ \hline
    Ave. MMSE $\alpha_{1,2}$            & 0.201 (0.013) &  0.202 (0.013)    & 0.2 \\ \hline
   Ave. Posterior Stdev. $\alpha_{1,2}$     & 0.053 (0.002) &  0.052 (0.002)    & -\\ \hline\hline
\end{tabular}

\caption{{\textbf{Sampler Analysis -}} Algorithm 1 is the pretuned
mixture proposal of Global ML move and local pretuned MCMC move;
Algorithm 2 is the Global adaptively learnt MCMC proposal.
Averages and a standard error are taken for the Bayesian point
estimators over 20 data sets, the standard errors are presented in
brackets $(\cdot)$. Note in all simulations the initial Markov
chain is started very far away from the true parameter values.} }}
\centering} \label{tab:Results1}
\end{table}

\newpage
\begin{table}
{\centering {\small{
\begin{tabular}{|c|c|c|}
    \hline
    \textbf{Model Rank}  & \textbf{Bayes Factors} \\
 \hline
    $r = 0$ & 3 (0.84) \\ \hline
    $\bf{r = 1}$ & 16 (0.93) \\ \hline
   $r = 2$ & 2 (0.92) \\ \hline
    $r = 3$ & 0 (-) \\ \hline
    $r = 4$ & 0 (-) \\ \hline\hline
    $r = 0$ & 0 (-) \\ \hline
    $r = 1$ & 5 (0.89) \\ \hline
   $\bf{r = 2}$ & 13 (0.91) \\ \hline
    $r = 3$ & 0 (-) \\ \hline
    $r = 4$ & 2 (0.92) \\ \hline\hline
    $r = 0$ & 0 (-) \\ \hline
    $r = 1$ & 0 (-) \\ \hline
   $r = 2$ & 4 (0.89) \\ \hline
    $\bf{r = 3}$ & 6 (0.90) \\ \hline
    $r = 4$ & 10 (0.94) \\ \hline\hline
    $r = 0$ & 0 (-) \\ \hline
    $r = 1$ & 0 (-) \\ \hline
   $r = 2$ & 0 (-) \\ \hline
    $r = 3$ & 2 (0.87) \\ \hline
    $\bf{r = 4}$ & 18 (0.89) \\ \hline\hline
\end{tabular}

\caption{{\textbf{Between Model Analysis -}} The true model rank
used to generate the data is presented in bold. TDMCMC is the
Trans-dimensional Markov chain Monte Carlo algorithm utilizing
adaptive MH within model moves and the global Independent between
model moves. The results represent the total number of times a
given rank is selected as the MAP estimate out of the 20
independent data sets, each of length T=100, analyzed.
Additionally, the average posterior model probability for these
cases is presented in brackets.} }} \centering}
\label{tab:Results2}
\end{table}

\newpage
\begin{landscape}
\begin{table}
{\centering {\small{
\begin{tabular}{|c|c|c|c|c|c|c|c|c|}
    \hline
    \textbf{Rank $\setminus$ T}   & \textbf{50} & \textbf{100} & \textbf{150} & \textbf{200} & \textbf{250} & \textbf{300} & \textbf{350} & \textbf{400} \\ \hline
    $r = 0$ & 0                 & 0                 & 0                 & 0                     & 0         & 0         & 0         & 0 \\ \hline
    $r = 1$ & 8.09 (0.78) &  3.77 (0.29)    & 7.01 (0.50)   & 11.51 (0.90)      & 1.69 (0.54)   & 2.71 (3.55)   & 3.14 (1.11)   & 7.77 (0.83)\\ \hline
    $r = 2$ & 2.91 (1.24)   &  2.33 (1.26)  & 4.61 (0.63)   & 25.36 (7.19)      & -5.33 (1.17)      &  -5.80 (0.97)     & 4.92 (1.06)   & -3.88 (1.11)\\ \hline
    $r = 3$ & -26.03 (1.06)     &  -8.45 (0.27)     & -37.25 (1.08)     & -55.79 (1.70)     & -14.61 (0.03)     &  -62.60 (3.31)    & $8.88$ ($0.88$) $\times 10^{-3}$  & -2.06 ($2.48 \times 10^{-2}$)\\ \hline
\end{tabular}

\caption{{\textbf{Log Bayes Factors:}} Analysis of VAR series of
US mini indexes as a function of data size. Average log Bayes
Factors and standard deviation of log Bayes Factors over 20
independent Markov chains each of chain length 20,000.} }}
\centering} \label{tab:Results2}
\end{table}

\begin{table}
{\centering {\small{
\begin{tabular}{|c|c|c|c|c|c|c|c|c|}
    \hline
    \textbf{Rank $\setminus$ T}   & \textbf{50} & \textbf{100} & \textbf{150} & \textbf{200} & \textbf{250} & \textbf{300} & \textbf{350} & \textbf{400} \\ \hline
    $r = 0$ & 0                     & 0                 & 0                 & 0 & 0         & 0         & 0         & 0 \\ \hline
    $r = 1$ & 4.81 (1.00)   & 3.14 (0.43)   & 5.36 (1.06)   & 5.92 (0.84)   & 3.32 (0.75)   & 1.30 (0.37)   & 7.30 (0.59)   & 3.10 (0.48)\\ \hline
    $r = 2$ &  -1.67 (12.78)    & 3.66 (3.87)   & -3.75 (3.04)  & -1.83 (2.61)      & -6.02 (3.22)      & 0.14 (6.51)   & -2.93 (1.96)  & -7.73 (2.46)\\ \hline
    $r = 3$ &  -42.44 (12.38)   & -48.58 (2.85)     & -33.12 (0.14)     & -100.42 (4.82)    & $-25.91 (6.52 \times 10^{-2})$    & -10.33 (0.72)     & -142.89 (3.31)    & -195.47 (4.71)\\ \hline
\end{tabular}
\caption{{\textbf{Log Bayes Factors:}} Analysis of VAR series of US Bonds (5,10,30 Year Notes) as a function of data size. Average log Bayes Factors and standard deviation over 20 independent Markov chains each of chain length 20,000.}
}}
\centering}
\label{tab:Results2}
\end{table}
\end{landscape}

\begin{figure}
\label{TracePlot}
\centering
\includegraphics[height=0.4\textheight,width=0.8\textwidth]{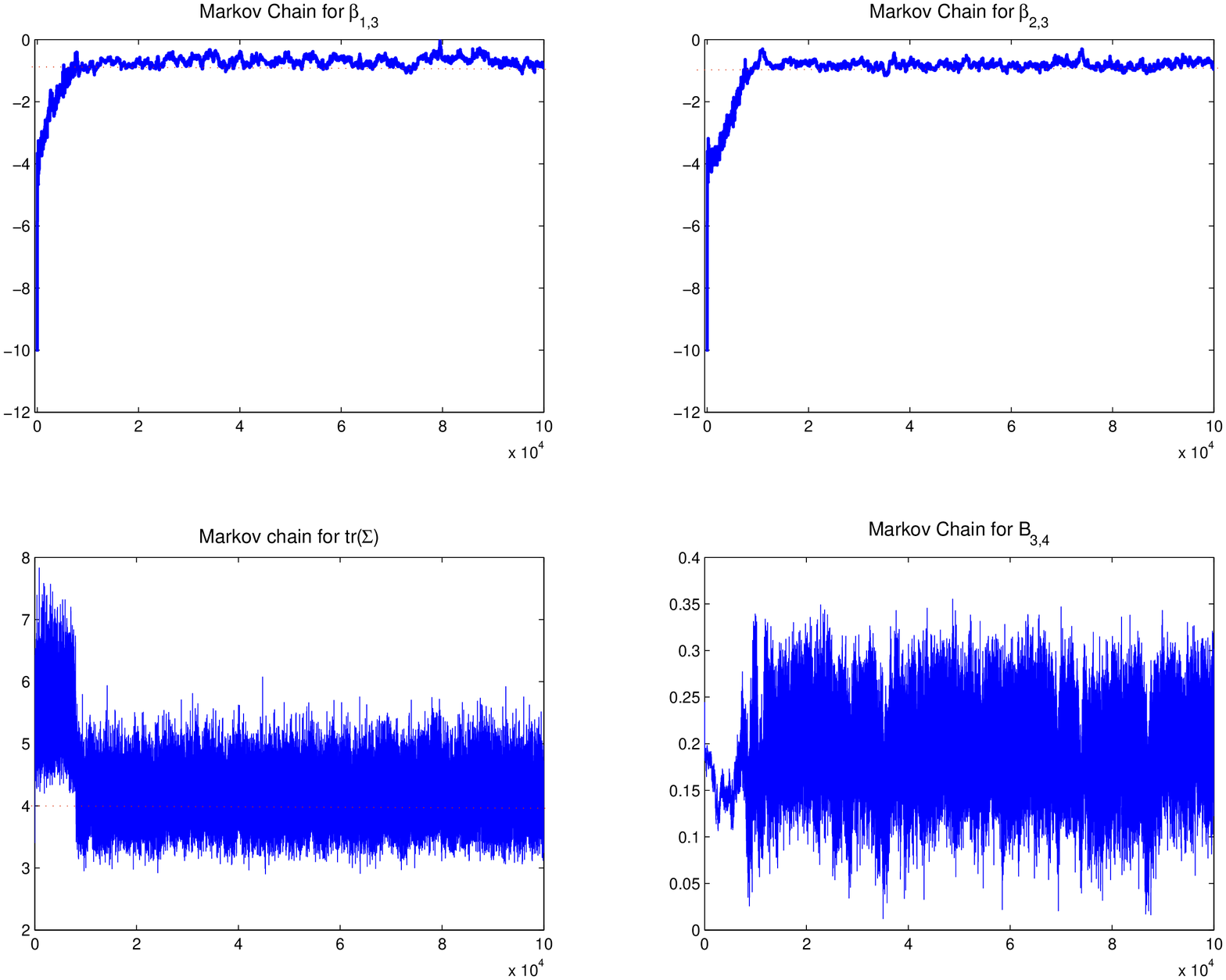}
\caption{Sample paths for posterior parameters, using 100 data points, true rank of $r=2$ known and an adaptive MCMC algorithm.}
\end{figure}

\begin{figure}
\label{TracePlot2}
\centering
\includegraphics[height=0.4\textheight,width=0.8\textwidth]{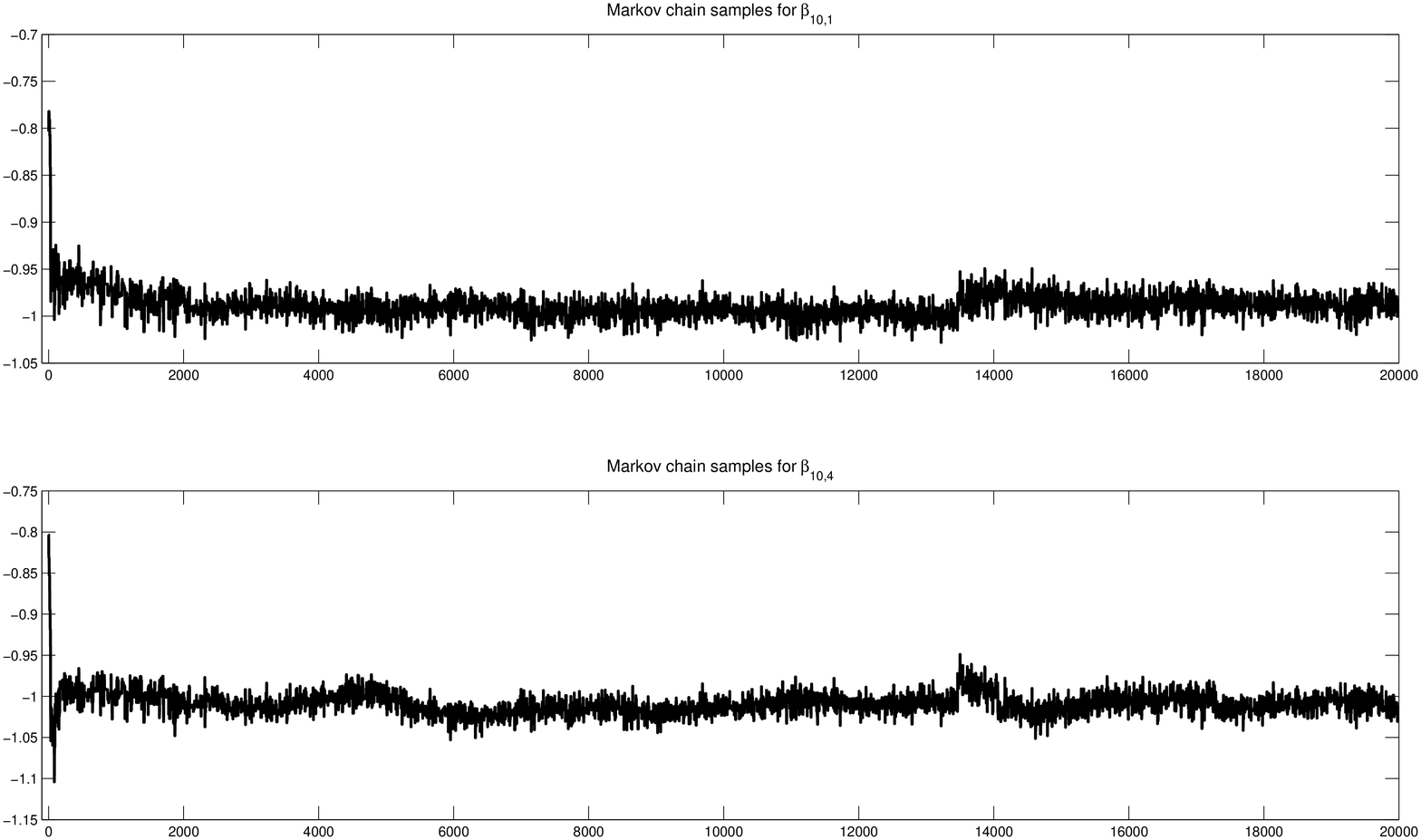}
\caption{Sample paths for posterior parameters, using 100 data points, true rank of $r=5$ known and an adaptive MCMC algorithm.}
\end{figure}

\begin{figure}
\label{IndexMini}
\centering
\includegraphics[height=0.4\textheight,width=0.8\textwidth]{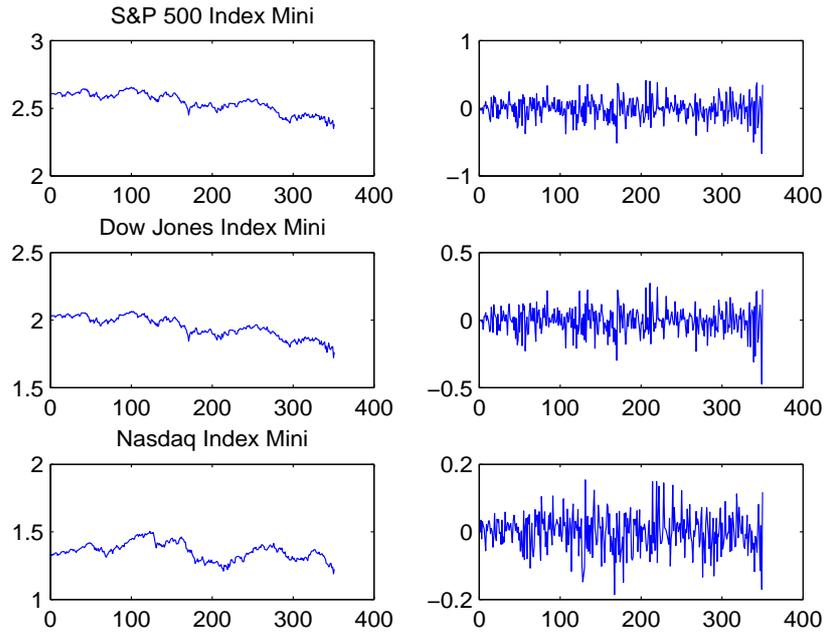}
\caption{S\&P 500, Dow Jones and Nasdaq mini Index daily close price data between 01-May-08 to 18-Sep-08. Left column plots represent scaled raw prices; Right plots represent difference data series.}
\end{figure}

\begin{figure}
\label{Bonds}
\centering
\includegraphics[height=0.4\textheight,width=0.8\textwidth]{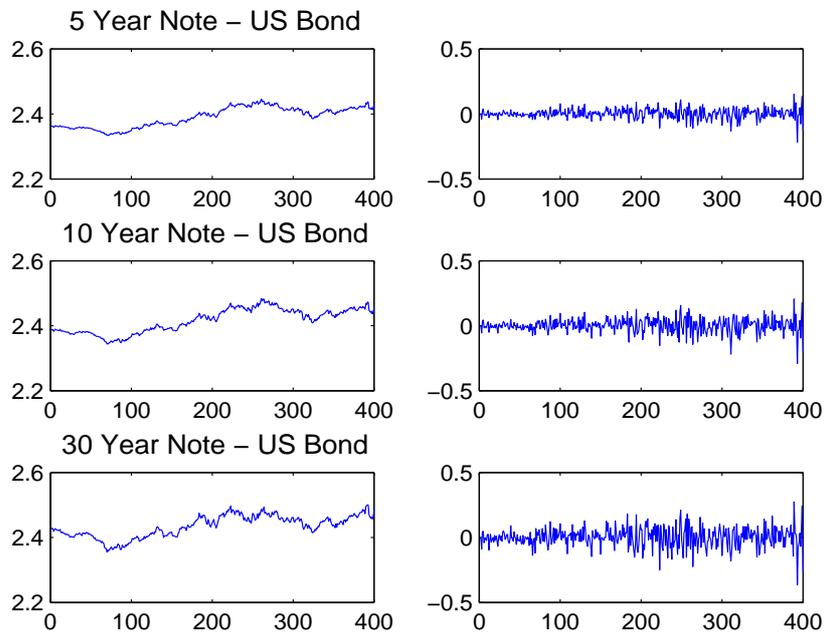}
\caption{5, 10, 30 Year Notes - daily close price data between 01-May-08 to 18-Sep-08. Left column plots represent scaled raw prices; Right plots represent difference data series.}
\end{figure}

\begin{figure}
\label{PredictionError}
\centering
\includegraphics[height=0.4\textheight,width=0.8\textwidth]{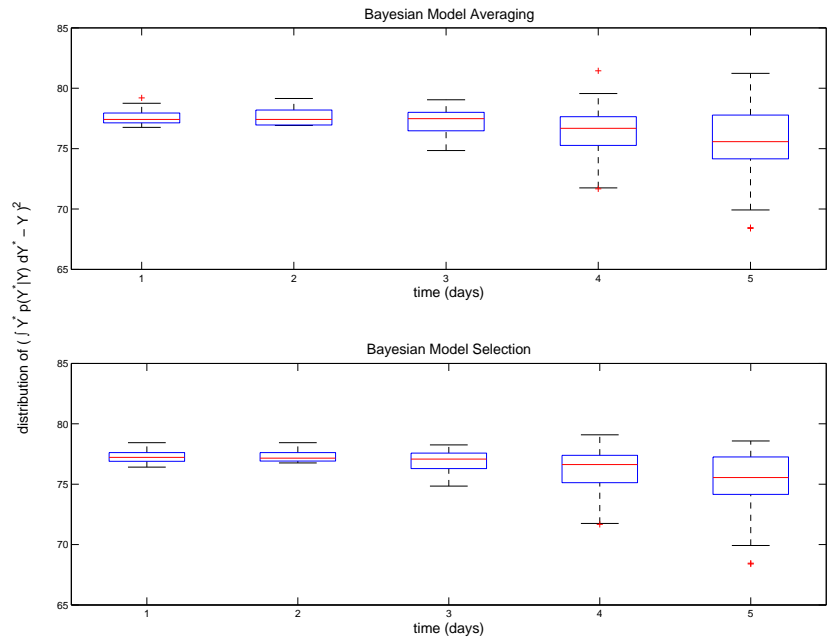}
\caption{Empirical distribution of the Bayesian Model Averaging
and Bayesian Model Order Selection, predictive performance for a
combination of mini-index and bond data, taken over random
intervals.}
\end{figure}

\end{document}